\documentclass[fleqn,usenatbib]{template/mnras}

\pdfoutput=1
\usepackage[T1]{fontenc}      % make sure font is supported
\usepackage[final]{microtype} % make sure font is supported
\usepackage{lmodern}          % use a modern font with T1 support

\usepackage{graphicx} % Including figure files
\usepackage{amsmath}  % Advanced maths commands
\usepackage{amssymb}  % Extra maths symbols
\usepackage{pifont}   % Extra maths symbols
\usepackage{xspace}   % handling of spaces after shortcuts

\usepackage{hyperref}   % Hyperlinks
\hypersetup{
  colorlinks=true,
  linkcolor=blue,
  citecolor=blue,
  filecolor=blue,
  urlcolor=blue,
  pdfauthor={Kueng, Rafael et al.},
  pdftitle={Models of gravitational lens candidates from Space Warps CFHTLS}
}

\newcommand*{\rot}{\rotatebox{90}}
\newcommand*{\OK}{\ding{51}}
\newcommand*{\NO}{\ding{55}}
\newcommand*{\UK}{\ensuremath{\text{--}}}

\newcommand{\SW}{Space\,Warps\xspace}
\newcommand{\SpL}{SpaghettiLens\xspace}
\newcommand{\asw}[1]{ASW000\,#1\xspace}
\newcommand{\sw}[1]{SW\,#1\xspace}

\newcommand{\figref}[1]{\ref{fig:#1}}

\newcommand{\Mstel}{\ensuremath{M_\text{stel}}}
\newcommand{\Mhalo}{\ensuremath{M_\text{halo}}}
\newcommand{\Msun}{\ensuremath{M_\odot}}

\newcommand{\haloindex}{\mathcal{H}}
\newcommand{\ER}{\ensuremath{r_\text{E}}\xspace} % einstein radius

\newcommand{\sqdeg}{\ensuremath{\,\text{deg}^2}\xspace}
\renewcommand{\arcsec}{\ensuremath{\,^{\prime\prime}}\xspace}

\def\pwidth{.32\linewidth}

\newcommand{\inclAC}[1]{\includegraphics[width=\pwidth]{img/#1/SW02_ASW000619d_011489_#1}}

\newcommand{\inclAF}[1]{\includegraphics[width=\pwidth]{img/#1/SW05_ASW0007k4r_012771_#1}}

\newcommand{\inclAJ}[1]{\includegraphics[width=\pwidth]{img/#1/SW09_ASW0002asp_5EKMWWVJHL_#1}}

\newcommand{\inclBJ}[1]{\includegraphics[width=\pwidth]{img/#1/SW19_ASW0001ld7_OS3CYAKLRT_#1}}

\newcommand{\inclCI}[1]{\includegraphics[width=\pwidth]{img/#1/SW28_ASW0007xrs_JHC3J2HYV7_#1}}
\newcommand{\inclCJ}[1]{\includegraphics[width=\pwidth]{img/#1/SW29_ASW0008qsm_TOFS7JNGEK_#1}}

\newcommand{\inclEC}[1]{\includegraphics[width=\pwidth]{img/#1/SW42_ASW00096rm_PQZR2WYE7X_#1}}

\newcommand{\inclFH}[1]{\includegraphics[width=\pwidth]{img/#1/SW57_ASW0008pag_5SXGXQYY6V_#1}}
\newcommand{\inclFI}[1]{\includegraphics[width=\pwidth]{img/#1/SW58_ASW0007iwp_4XBJWT3COV_#1}}

\newcommand{\inclGrid}[1]{ %
\inclFI{#1} \inclCI{#1} \inclFH{#1}
\inclAF{#1} \inclEC{#1} \inclBJ{#1}
\inclAJ{#1} \inclCJ{#1} \inclAC{#1}
}

\title[Lens models for Space Warps CFHTLS]{Models of gravitational
    lens candidates from Space Warps CFHTLS}

\author[K\"ung et al]{Rafael K\"ung,$^{1}$
Prasenjit Saha,$^{1}$
Ignacio Ferreras,$^{2}$
Elisabeth Baeten,$^{3}$
\newauthor
Jonathan Coles,$^{4}$
Claude Cornen,$^{3}$
Christine Macmillan,$^{3}$
Phil Marshall,$^{5}$ 
\newauthor
Anupreeta More,$^{6}$
Lucy Oswald$^{7}$
Aprajita Verma$^{8}$
and Julianne K. Wilcox$^{3}$
\\
$^{1}$Physik-Institut, University of Zurich, Winterthurerstrasse 190, 8057 Zurich, Switzerland\\
$^{2}$Mullard Space Science Laboratory, University College London, Holmbury St Mary, Dorking, Surrey RH5 6NT, UK\\
$^{3}$Zooniverse, c/o Astrophysics Department, University of Oxford, Oxford OX1 3RH, UK \\
$^{4}$Physik-Department, Technische Universit\"at M\"unchen
James-Franck-Str.~1, 85748 Garching, Germany\\
$^{5}$Kavli Institute for Particle Astrophysics and Cosmology, Stanford University, 452 Lomita Mall, Stanford, CA 94035, USA\\
$^{6}$Kavli IPMU (WPI), UTIAS, University of Tokyo, Kashiwa, Chiba 277-8583, Japan\\
$^{7}$Murray Edwards College, University of Cambridge, Cambridge CB3 0DF, UK\\
$^{8}$Sub-department of Astrophysics, University of Oxford, Denys Wilkinson Building, Keble Road, Oxford, OX1 3RH, UK\\
}

\date{Accepted 2017 November 19. Received 2017 November 19; in original form 2017 May 02}

\pubyear{2017}

\begin{document}
\label{firstpage}
\pagerange{\pageref{firstpage}--\pageref{lastpage}}
\maketitle

\begin{abstract}
We report modelling follow-up of recently-discovered
gravitational-lens candidates in the Canada France Hawaii Telescope
Legacy Survey. Lens modelling was done by a small group of
specially-interested volunteers from the \SW citizen-science community
who originally found the candidate lenses.  Models are categorised
according to seven diagnostics indicating (a)~the image morphology and
how clear or indistinct it is, (b)~whether the mass map and synthetic
lensed image appear to be plausible, and (c)~how the lens-model mass
compares with the stellar mass and the abundance-matched halo mass.
The lensing masses range from $\sim10^{11}M_\odot$ to
$>10^{13}M_\odot$. Preliminary estimates of the stellar masses show a
smaller spread in stellar mass (except for two lenses): a factor of a
few below or above $\sim10^{11}M_\odot$.  Therefore, we expect the
stellar-to-total mass fraction to decline sharply as lensing mass
increases.  The most massive system with a convincing model is
J1434+522 (\sw{05}).  The two low-mass outliers are J0206$-$095
(\sw{19}) and J2217+015 (\sw{42}); if these two are indeed lenses,
they probe an interesting regime of very low star-formation
efficiency.  Some improvements to the modelling software
(SpaghettiLens), and discussion of strategies regarding scaling to
future surveys with more and frequent discoveries, are included.
\end{abstract}


%%%%%%%%%%%%%%%%%%%%%%%%%%%%%%%%%%%%%%%%%%%%%%%%%%%%%%%%%%%%%%%%%%%%%%%
\section{Introduction}

By a curious coincidence, the typical escape velocity of massive
galaxies -- of order a few hundred km/s -- is such that $v_{\rm
  esc}^2/c^2$, expressed as an angular distance, is comparable to the
apparent sizes of galaxies at cosmological distances.  This
coincidence is fortunate, because it makes the gravitational lensing
deflection angle of distant galaxies ($\alpha\sim 2v_{\rm esc}^2/c^2$)
comparable to their size on the sky. As a result, strong lensing by
galaxies produces images that probe their host dark matter halos,
providing a useful tool to understand galaxy formation.  While there
is a general consensus about the basic mechanisms at play, involving
gravitational collapse, fragmentation, and mergers of dark-matter
clumps, into which gas fell, cooling through radiative processes to
form dense clouds and eventually stars, there is much debate about the
details \citep[for a summary, see][]{2012RAA....12..917S}.  In
particular, the nature of dark matter remains mysterious: most
researchers take it to be a collisionless non-relativistic fluid (cold
dark matter or CDM) readily studied by simulations \citep[for example,
  the influential Millennium
  simulation,][]{2005Natur.435..629S}. However, alternative scenarios,
where dark matter has exotic dynamical properties
\citep{2010MNRAS.405...77S,2016ApJ...818...89S}, or is not really
matter at all, but a modification of gravity
\citep{2016PhRvL.117t1101M}, have also been considered.

All this motivates the use of strong gravitational lensing over galaxy
scales to study the mutual dynamics of dark matter, gas and stars.
Several studies in recent years have done so \citep[see,
  e.g.,][]{2009ApJ...703L..51K,2011ApJ...740...97L,2012MNRAS.424..104L,
  2016MNRAS.459.3677L,2016MNRAS.456..870B} but it is desirable to
enlarge the samples from tens of lensing galaxies to thousands.  Doing
so requires both finding more lenses and also modelling their mass
distribution.  Recent searches through the CFHT Lens Survey
\citep[CFHTLS,][]{2012MNRAS.427..146H} using arc-finders
\citep[e.g.,][]{2012ApJ...749...38M,2014A&A...567A.111M,2014ApJ...785..144G,2017arXiv170401585S}
by either machine learning methods
\citep[e.g.,][]{2016A&A...592A..75P,2017arXiv170302642L} or visual
inspection by citizen-science volunteers through the \SW project
\citep{2016MNRAS.455.1191M} have, between them, discovered an average
of four lenses per square degree, so one can be optimistic about
finding many thousands of lenses in the next generation of wide-field
surveys, from ground-based surveys such as LSST in the optical window
and SKA in radio, to space-based missions such as Euclid and WFIRST
\citep{2010MNRAS.405.2579O,2015ApJ...811...20C}.  The availability of
the recently published year~1 result of the Dark Energy Survey
\citep{2017arXiv170801531D}, which covered 1800\sqdeg, suggests that
the number of lenses could grow by at least an order of magnitude
within a very short time period.
% Moreover, ongoing work by Denzel et al. (in prep) will provide much more
% strict bounds on the uncertainties presented in this work.

The expected flood of new lens discoveries will need a similarly large
modelling effort to reconstruct their mass distributions.
Lens-modelling robots have started to be developed
\citep{2017arXiv170807377N,2017arXiv170808842H} but so far are able to
handle only very clean systems.  For typical lenses, human input is
still needed.  With that in mind \cite{2015MNRAS.447.2170K} developed
a new modelling strategy, implemented as the SpaghettiLens\footnote{\url{http://labs.spacewarps.org/spaghetti/}}
system.
The idea is to collaborate with experienced members of the
citizen-science community, who have already participated in lens
discovery through \SW, as well as several other projects involving
astronomical data.  The system was tested on a sample of simulated
lenses, which were part of the training and testing set in \SW.

This paper follows up that study by applying SpaghettiLens to
candidates discovered through \SW.  We present results from
the modelling of 58 of the 59 lens candidates reported by
\cite{2016MNRAS.455.1191M}.  Each lens candidate was modelled
following a collaborative refinement process, where anyone interested
could improve the analysis by modifying an existing model or creating
a new one. Note the difference with respect to the main \SW 
project, where volunteers from a group of $\gtrsim10^4$ people make
independent contributions.  Each person is presented with a random
selection of survey-patches and invited to (in effect) vote on each.
The system estimates the skill level of each volunteer according to
test-patches interspersed with the real data, and weights their votes
accordingly \citep{2016MNRAS.455.1171M}.  There is an active forum for
volunteers, but since everyone is seeing different data samples with
minimal overlap, the forum has little if any influence on the votes.
In SpaghettiLens, the number of volunteers is significantly lower, but
the level of interaction is higher.  The resulting model represents a
consensus among contributors, as to the best that could be achieved
with the available data and software.

We emphasize that the interpretation of the results presented here is
tentative, because the systems are lens candidates at this stage, not
secure lenses.  Moreover the candidate-lens redshifts have large
uncertainties, while the candidate-source redshifts can only be
guessed at present.  Nevertheless it is interesting to explore the
trends observed with the already available data, based on some 300 models
created by 8 volunteers for 58 candidate lenses found in the 150\sqdeg
CFHTLS.

This paper is organized as follows:
Section~\ref{sec:candidates_models} introduces the candidate lenses,
their models and the diagnostics applied to them.  The following
sections elaborate on the diagnostics.  Image morphology diagnostics
are explained in Section~\ref{sec:morph} and diagnostics based on the
mass models are discussed in Section~\ref{sec:massmodels}.  Stellar masses are
presented in Section~\ref{sec:stellar-mass}, to compare stellar and
lensing masses.  Finally, section~\ref{sec:summary}
summarises and tabulates the diagnostics in Table~\ref{tab:models}.

We include three appendices devoted to various technical issues
relating to the modelling.  The online supplement gives the results
from all the modelled systems generated for all the lensing
candidates.

%%%%%%%%%%%%%%%%%%%%%%%%%%%%%%%%%%%%%%%%%%%%%%%%%%%%%%%%%%%%%%%%%%%%%%%
\section{The candidates and models}
\label{sec:candidates_models}

\SW is a citizen science gravitational lens search
\citep{2016MNRAS.455.1171M}.  Its first run searched the CFHT Legacy
Survey, a survey carried out in five optical bands
($u^*$,$g^\prime$,$r^\prime$,$i^\prime$,$z^\prime$) covering a total
area of 160\sqdeg divided up into four fields W1 -- W4.  The
MegaPrime camera, used for the survey, has a field of view of
1\sqdeg, with $0.186\arcsec$ pixels. The flux limit of the
survey is $g^\prime<25.6\,\text{AB}$ ($5 \sigma$) with a typical seeing
FWHM $\sim 0.7\arcsec$ \citep{2013MNRAS.433.2545E}. The cutouts
shown to the volunteers were colour composites of the $g^\prime$,
$r^\prime$ and $i^\prime$ bands from randomly chosen regions.
The programme resulted in 59 new lens candidates, of which 29 are
considered high quality.
Moreover, 82 previously known lens candidates were ``rediscovered''.
The candidates have broadly similar redshifts ($0.2<z<1.0$), Einstein
radii between 0.7 and 5\arcsec, and arc fluxes in $g$-band AB
magnitudes between 22 and 26. These properties are
similar to those found by previous robotic searches like ArcFinder and
RingFinder \citep{2016MNRAS.455.1191M}.

\begin{figure}
  \includegraphics[width=\linewidth]{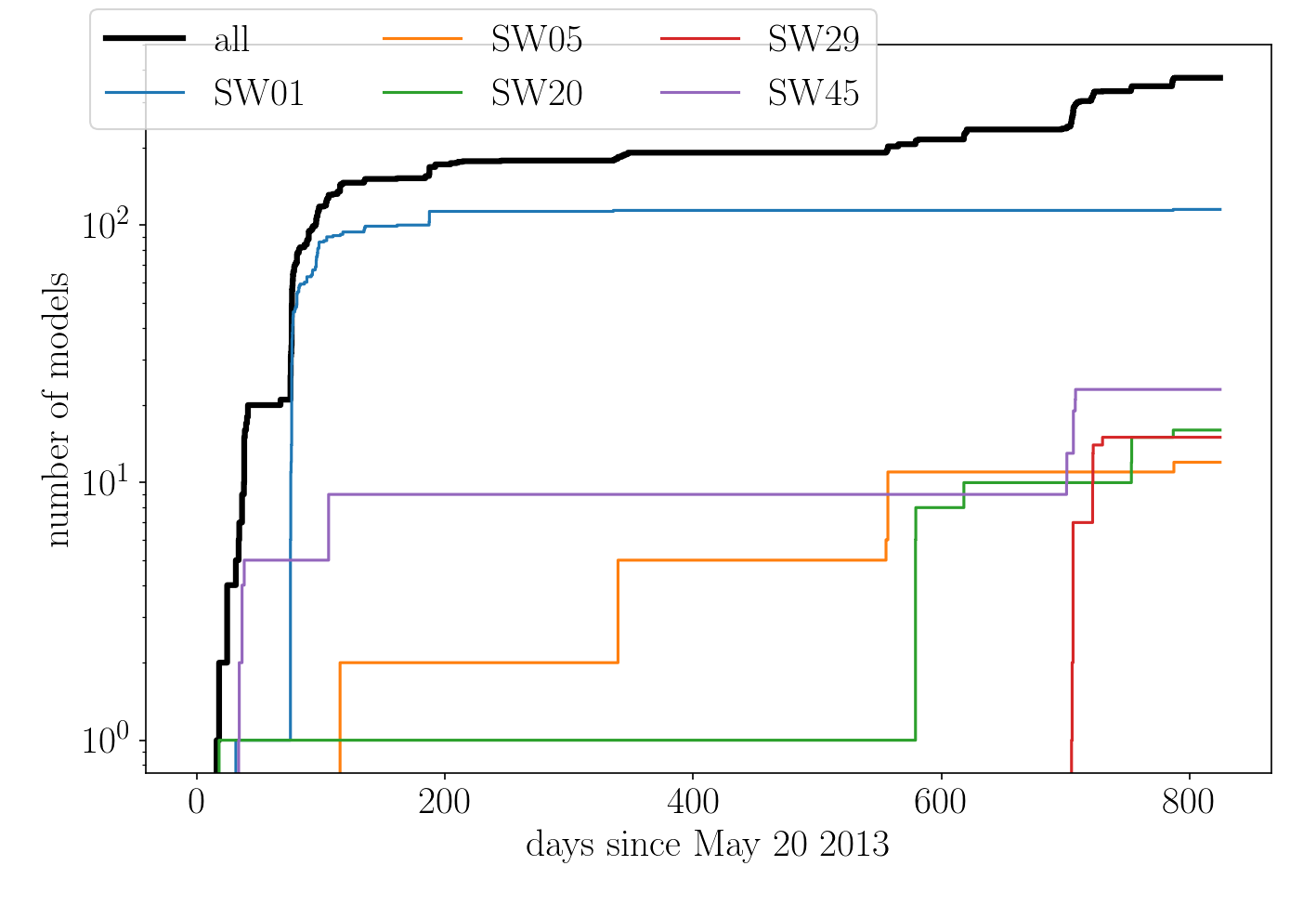}
  \caption{ Number of models generated over time in days since the
    launch of \SpL; the black line shows all models, the coloured
    lines show the five most active candidates. A total of 377 models
    were generated for 58 lens candidates, eight users contributed
    more than five models per person. }
  \label{fig:time}
\end{figure}

A first version of \SpL was launched in May 2013, while
the first run of \SW was ongoing.  Fig.~\ref{fig:time} shows the
total number of models generated since launch for our lens candidates.
A small number of volunteers started creating models of possible lens
candidates from \SW that were debated in the discussion forum.  In the
beginning of August 2013 (around day 70 since release) a modelling
challenge for the candidate later to be identified as \sw{01} was
launched.  A group of five volunteers presented their results in an
online letter
converging on a set of 30 models. In April 2015 (day 700),
the first major version of \SpL was released. At the same time,
the list of identified candidates was made available \citep[as a preprint
  of][]{2016MNRAS.455.1191M}.  At this stage, the volunteers were
asked to create models for all the lens candidates that were still
missing.  The work done for candidate \sw{29} show that the
volunteers converged on a favourable model much faster, after 15
models, generated in 30 days.  The total modelling effort resulted in
a set of 377 \SpL models for all but one of the 59 \SW lens
candidates; candidate \sw{39} was not modelled by any volunteer.  Over
its runtime, the system logged eight major contributors (defined as
volunteers that generated, at least, five models).

In parallel, \citet{2015MNRAS.447.2170K} tested the system on
simulated lenses and identified some areas for improvement.  In
\S~\ref{subsec:sourcefit} we introduce fitting of the surface
brightness profile of the source.  This feature has not been included
yet in SpaghettiLens, but we carried the analysis during
post-processing for a few interesting cases.  In \S~\ref{subsec:hires}
we show that fine-grained mass maps within the central region
relieves a tendency in the earlier work for mass profiles to be too
shallow. In \S~\ref{subsec:parameter} we consider the possibility
of fitting a parametric lens model to the ensemble.

We characterise each model with seven diagnostics, grouped into three
categories, whose purpose is to help identify the most likely cases of
a gravitational lens, as well as flag the most interesting candidates
for future follow-up observations. The diagnostics are as follows.

\begin{itemize}
\item Firstly we have diagnostics based on morphology.
  Section~\ref{sec:morph} and Fig.~\ref{fig:splinput} explain
  this diagnostic in more detail. We consider:
\begin{enumerate}
\item Whether the images are unblended:  Distinct, unblended images are
  an advantage in modelling, although not essential.
\item Whether all images are discernible.  The topography of an
  arrival-time surface, as encoded by a spaghetti diagram, may require
  more images than are visible, in which case the modeller has to
  insert conjectural image positions.
\item Whether the lens is isolated.
\item The image morphology concisely described: double or quads,
  further sub-categorised to indicate the elongation direction of the
  lensing mass.
\end{enumerate}
\item Secondly we have mass models, covered in Section~\ref{sec:massmodels}.
  We assess the following points:
\begin{enumerate}
\setcounter{enumi}{4}
\item Whether the mass map is reasonable (see Fig.~\ref{fig:kappa}).
\item Whether the arrival-time surface and synthetic image are
  plausible. In particular, an unsatisfactory model is flagged when 
  additional images are implied in regions
  where they are not observed (see Figs.~\ref{fig:arriv}, \ref{fig:synth} \ref{fig:encl}).
\end{enumerate}
\item Thirdly, whether the total implied lensing mass determined from
  the lens model is plausible, given the photometry of the lensing
  galaxy.  In this work we define the lensing mass as the sum of all
  mass tiles in the model.  These mass tiles reach out to typically
  twice the radius of the outermost image.
  Section~\ref{sec:stellar-mass} explains how we compare the lensing
  mass with the stellar mass.  Galaxies may have halos extending well
  beyond the mass maps in the models, so the calculated lensing mass
  is only a lower bound on the total galaxy mass.  Hence, the lensing
  mass should be somewhere between the stellar mass (lower bound) and
  the total halo mass (upper bound).  Extrapolation of the lens model
  to estimate the full virial mass may be possible
  \citep[cf.][]{2012MNRAS.424..104L}, but is not attempted in this
  work.
\begin{enumerate}
\setcounter{enumi}{6}
\item We then introduce a so-called halo index ($\haloindex$), that
  quantifies how the lensing mass compares with these two
  constraints (see Fig.~\ref{fig:stelmass}).
\end{enumerate}
\end{itemize}

%%%%%%%%%%%%%%%%%%%%%%%%%%%%%%%%%%%%%%%%%%%%%%%%%%%%%%%%%%%%%%%%%%%%%%%
\section{Image morphology}
\label{sec:morph}

The main input of a modeller to the process is a markup of the
candidate lens system, which we call a spaghetti diagram.  This is a
sketch of the arrival-time surface from a point-like source, with
proposed locations of maxima, minima and saddle points, and an implied
time-ordering of the images.  Such information encodes a starting
proposal for the mass distribution.  A spaghetti diagram is thus a
completely abstract construction, and moreover it refers to a
simplified system with a point source.  However, spaghetti diagrams
are intuitive because they tend to resemble the form of lensed arcs
\citep[see Fig.3 of][]{2008MNRAS.383..857F}, and of course they are
simple to draw, and easy to vary and refine in an open collaborative
environment. This makes them very practical for non-professional lens
enthusiasts in the citizen-science community.  Details and tests are
given in \citet{2015MNRAS.447.2170K}.

\begin{figure*}
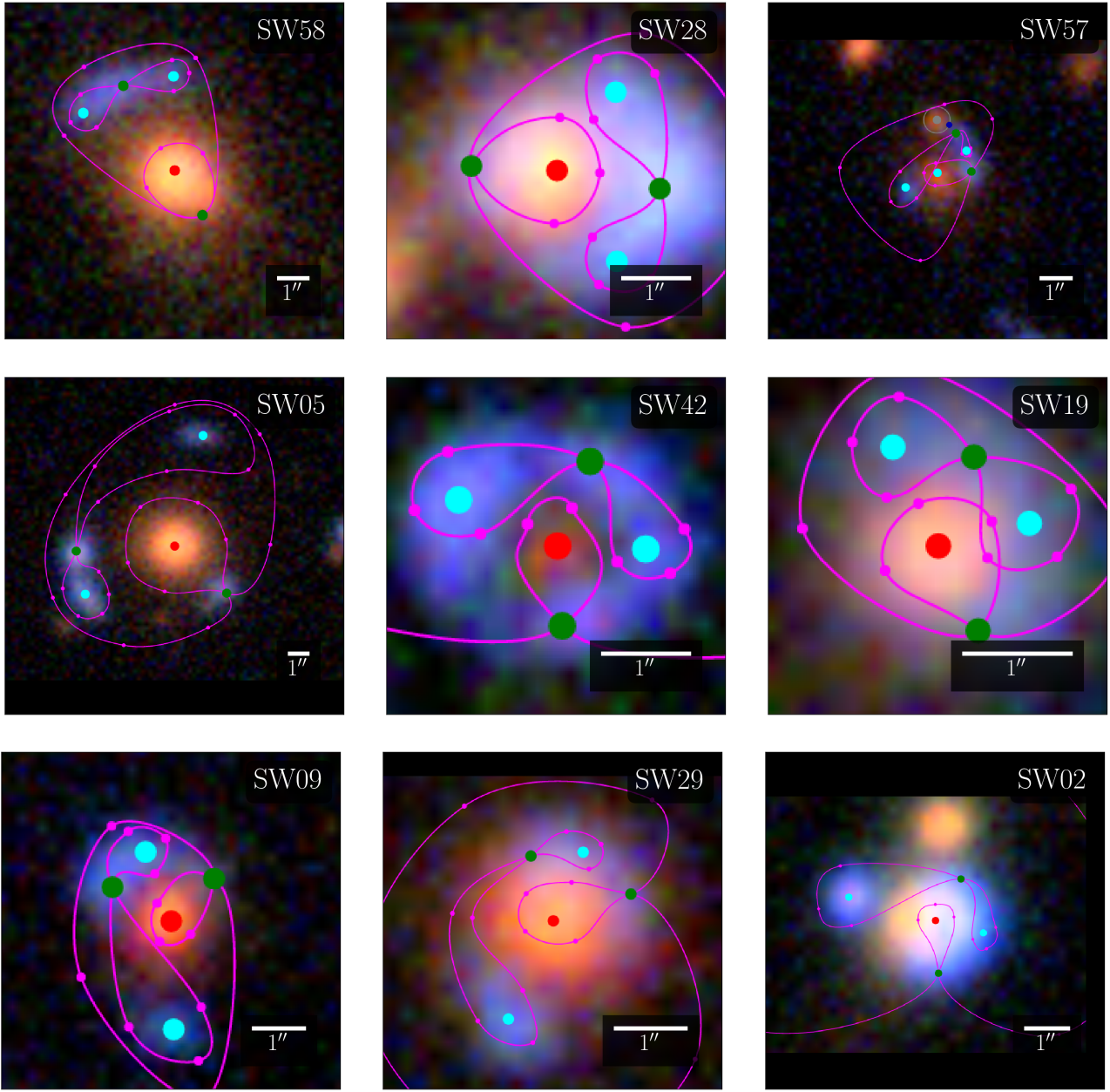

\inclGrid{spl-input}
\caption{Nine of the lens candidates marked up with spaghetti
  diagrams.  Red, blue and green dots are proposed locations for
  maxima, minima and saddle points of the arrival time respectively.
  The curves help guide the placement of the dots, but their precise
  appearance has no significance.  This images are screenshots from the 
  SpaghettiLens user interface, which applies interpolation to background
  images. The scaling is adjusted to fit the other images.
  This selection includes the
  best-modelled systems, but also one case (\sw{57} at upper right) of
  unsuccessful modelling.  Since the modelling process is
  collaborative among the volunteers, with anyone welcome to
  contribute new models or modify existing ones, there are variant
  spaghetti diagrams for all the modelled systems.  The online
  supplement displays all the models presented for discussion during
  this work.  
\label{fig:splinput}}
\end{figure*}

We now discuss the diagnostics that can be taken from the process of
drawing the spaghetti diagram, even before detailed mass-modelling
takes place.  Fig.~\ref{fig:splinput} shows nine examples, each
consisting of a cutout of the \SW  image of a lens candidate,
marked up with a spaghetti diagram.
The volunteers are initially presented a square image with side
82\arcsec (440~pixels) in full size, but they have the
ability to zoom in individually.
The cutouts presented in Figs.~\ref{fig:splinput} to \ref{fig:kappa}
are rescaled during post-processing, relative to the outermost image.

All the examples shown in Fig.~\ref{fig:splinput} identify five locations:
the centre of the main lensing galaxy, which is also a maximum of the
arrival time surface (red dots); two minima (blue dots); and two
saddle points (green dots and also self-intersections of the curves).
Only two (\sw{05} and \sw{42}) of the nine systems, however, have five
distinct features in plausible locations.  In the other seven
examples, an arc is interpreted as a blend of three or four images.
This defines the `unblended images' diagnostic.  Note that this
characterisation could be different if the spaghetti diagram were 
different.  For example, \sw{28} has also been modelled with the arc on
the right interpreted as a single image, rather than as three images
as shown in the figure; for such a model, the `unblended images'
diagnostic would be true.

The second diagnostic checks whether all images are visible.  For example,
we see in \sw{58} at the top left of Fig.~\ref{fig:splinput} that an
image at the second minimum is conjectural and does not correspond to
any visible feature.

The third diagnostic tests whether the lens is isolated, or whether
other galaxies could contribute to the lensing, based on visual inspection 
of the field.
For this, we do not
consider stars or other clearly foregound objects.  Additional
galaxies contributing to the lensing mass can be marked by the
volunteer alongside the spaghetti diagrams.  An example can be seen as
the grey dot and circle in the cutout with \sw{57} at the top right of
Fig.~\ref{fig:splinput}.  Objects marked in this way are modelled as
point masses.  Other possible contributors to lensing are galaxies or
groups that are not in the immediate vicinity of the lensed images,
yet massive enough to exert an influence. These are accounted for by
allowing a constant but adjustable external shear.

We remark that the lack of expected lensed images, or the presence of
blended images or perturbing galaxies do not imply that a given
candidate is unlikely to be a lens.  This means, rather, that the
models are more uncertain and perhaps could be more easily improved by
trying further variations in the markup.

The fourth diagnostic is based on the fact that the arrangement of
lensed images of a pointlike source through a non-circular
gravitational lens depends on the location of the source relative to
the long and short axes of the lens.  This dependence is quite robust
and independent of many other details of the lensing mass
distribution.  Sources close to being dead-centre behind a lens tend
to produce quads; sources at larger transverse distance tend to
produce doubles.  We denote these as Q and D, respectively, and add a
prefix to the Q systems, as follows: we write LQ if the source is
inferred as displaced along the long axis of the lens, SQ if displaced
along the short axis, IQ if inclined to both axes, CQ if only very little or
no displacement is evident.  Although the
unlensed source and its location are obviously not seen, the LQ, SQ, IQ and CQ
cases correspond to easily-seen image morphologies \citep[see,
  e.g.,][]{2003AJ....125.2769S}.
\begin{itemize}
\item The simplest is the LQ case: this creates a saddle point and two
  minima in an arc, with the second saddle point on the other side of
  the lensing galaxy, closer to the galaxy than the arc.  \sw{58}
  and \sw{28} in the upper row of Fig.~\ref{fig:splinput} are typical
  examples of LQ.  If the source were moved outwards along the short
  axis, the minimum-saddle-minimum set would merge into a single minimum,
  leaving a D system; the transition is known as a cusp catastrophe.
\item The middle row of Fig.~\ref{fig:splinput} shows three IQ
  systems: \sw{05}, \sw{42} and \sw{19}.  This type has a characteristic
  asymmetry, often with two images close together.  If the source
  were moved outwards, the minimum-saddle pair would merge and
  mutually cancel, leaving again a D system. This transition is known
  as a fold catastrophe.
\item The lower row of Fig.~\ref{fig:splinput} shows three SQ
  systems: \sw{09}, \sw{29} and \sw{02}. The failed model, \sw{57}, at top right
  may also belong to this category.  Here the images have a
  fairly symmetric arrangement with an arc and a counter-image on the
  other side, but the spaghetti diagram is completely asymmetric.  If
  the source were moved outwards along the long axis, the
  saddle-minimum-saddle set  would merge into a single saddle --- another
  form of cusp catastrophe.  SQ can be visually distinguished from LQ
  by the relative distances of the arc and the counter-image.  For
  SQ the arc is closer, for LQ the counter-image is closer.
\end{itemize}
CQ systems are often called `cross' or `Einstein-cross' systems; IQ
systems are sometimes called `folds', with `cusp' commonly used for
both SQ and LQ.  The labels `short-axis quad' and so on are not
standard in the literature, but the morphological classification they
express is familiar to experienced modellers.  Hence they can be
useful to researchers wishing to apply other modelling methods to the
same systems.

%%%%%%%%%%%%%%%%%%%%%%%%%%%%%%%%%%%%%%%%%%%%%%%%%%%%%%%%%%%%%%%%%%%%%%%
\section{Mass models}\label{sec:massmodels}

Once a spaghetti diagram has been drawn on a web browser, it is
forwarded to a server-side numerical framework, which searches for
mass maps consistent with the image locations, parities and time
ordering, given by the modeller.  The mass maps are made up of mass
tiles and are free-form, but are required to be concentrated around
the identified lens centre \citep[see][for the precise formulation of
  the search problem]{2014MNRAS.445.2181C}.  The modeller can also set
various options for the search, such as the number of mass tiles and
the extent of the mass map; all options have defaults.  Typically,
there are $\sim500$ mass tiles arranged in a disc, centred on the
lensing galaxy and extending to twice the radius of the outermost
image.  By default, the mass distributed is required to have
$180^\circ$ rotation symmetry, but this option can be unset.  Assuming
mass distributions can be found -- which in practice is usually the
case -- a statistical ensemble of 200 two-dimensional mass maps is
returned.  From each mass map, further quantities such as lens
potentials or enclosed masses can be derived.  Thus, there are
ensembles of 200 values for the surface density at any point, for the
enclosed mass within a given radius and so on.  In this paper we
present the averages and ranges of different quantities, but other
quantities such as 90\% confidence ranges could also be computed.  The
whole ensemble of mass maps, along with derived quantities and
uncertainties, makes up one SpaghettiLens model.

The mass will naturally depend on the lens and source redshifts, which
are unknown when a lens candidate is first identified.  However, this
is not a problem, because a model can be trivially rescaled to use
better redshift values, as they become available.  The mass
normalisation of the models is proportional, through the angular
critical density, to the factor $d_L\, d_S\, d_{LS}^{-1}$, where
$d_L,d_S$ and $d_{LS}$ are the usual angular-diameter distances.  In
the redshift ranges typical of galaxy lensing, the normalisation
factor is roughly proportional to the lens redshift, and weakly
dependent on the source redshift.  This work applies photometric
redshifts from the CFHTLS pipeline \citep{2009A&A...500..981C} to the
candidate lensing galaxies; the values range from $z=0.2$ to 1 (see
Table~\ref{tab:models}).  The source redshifts are assumed as $z=2$,
unless an unambiguous photometric redshift is available.  The lens
redshifts entail rather big uncertainties, up to a few tens of percent
\citep[see Fig.~5 in][]{2009A&A...500..981C}. The lensing masses would
be uncertain at the same level.  On the other hand, the lensing masses
in the sample range from $10^{11}M_\odot$ to $10^{13}M_\odot$; in
comparison, the redshift uncertainty is not so important in this
preliminary analysis.

The ensemble of mass maps can be post-processed in many different
ways.  Four different graphical quantities are particularly useful.

\begin{figure*}
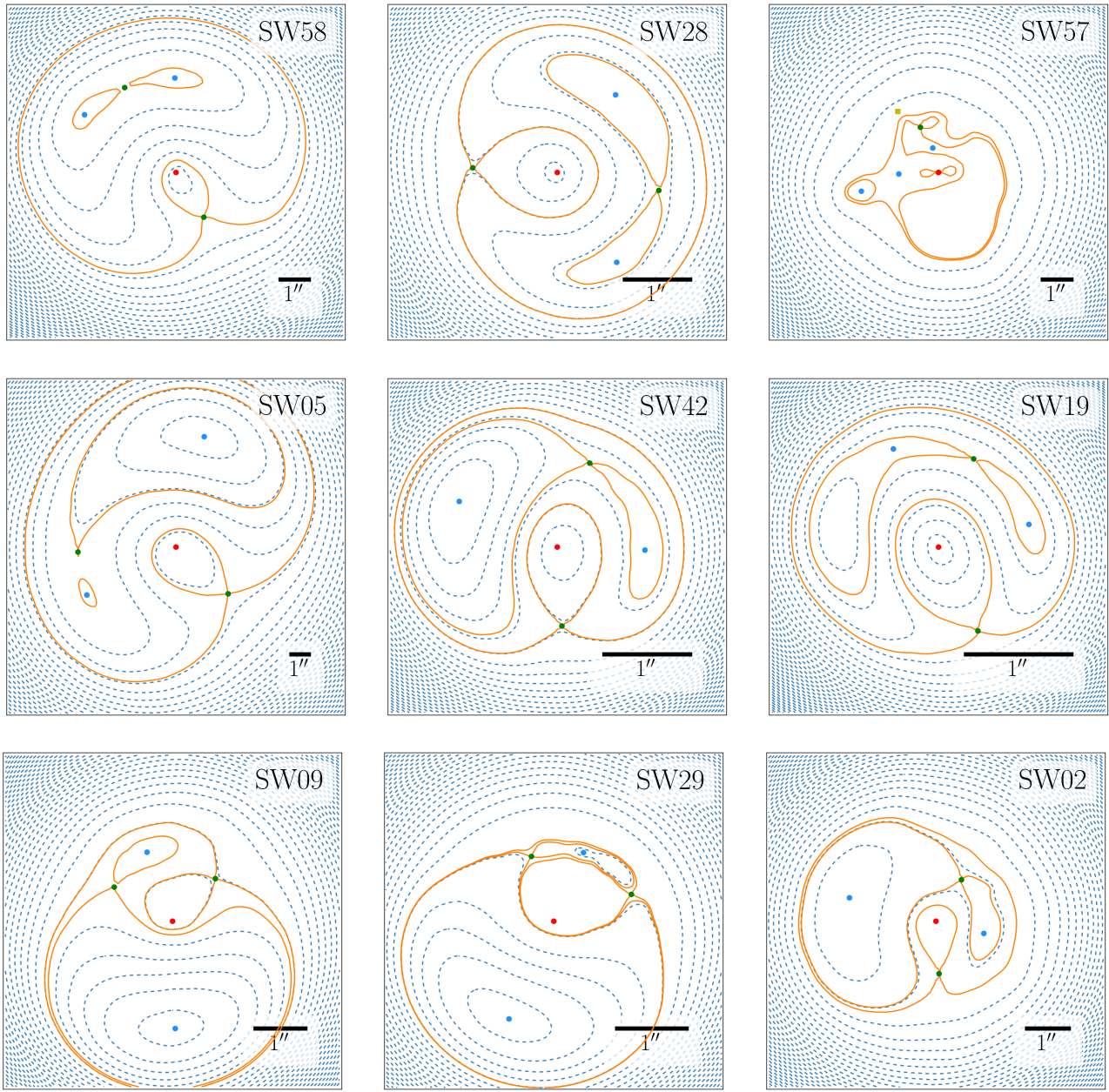

\inclGrid{arrival_spaghetti}
\caption{Arrival-time surfaces for models of the systems from
  Figure~\ref{fig:splinput}.  The registration differs slightly from
  Figure~\ref{fig:splinput}, but the coloured dots represent exactly
  the sky positions specified in the earlier figure.  The orange
  contours only qualitatively resemble the earlier pink curves, as
  they are now precise saddle-point contours from lens models.
  \label{fig:arriv}}
\end{figure*}

\begin{figure*}
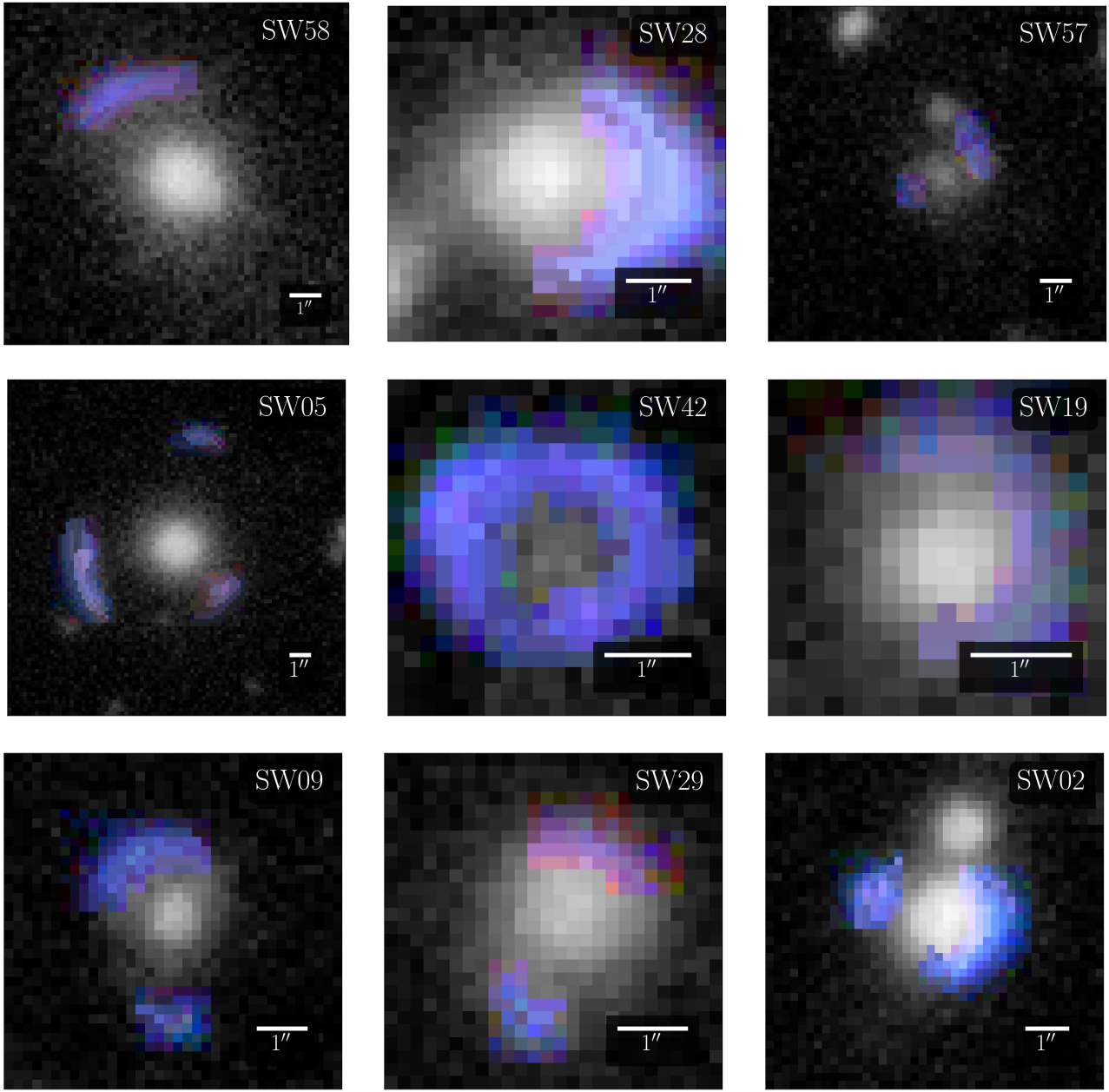

\inclGrid{nsynth}
\caption{Synthetic images of the systems from
  Figure~\ref{fig:splinput}, derived from the lens models.  The
  reconstructed lensed features keep the Space~Warps false-colour
  scheme from Figures~\ref{fig:splinput}.  The rest has been changed
  to black-and-white. 
  \label{fig:synth}}
\end{figure*}

Fig.~\ref{fig:arriv} shows the arrival-time surfaces corresponding to
the spaghetti diagrams of Fig.~\ref{fig:splinput}.  The arrival-time
contours look like machine-made elaborations of the input spaghetti
diagrams.  If the saddle-point contours in the arrival time are
qualitatively the same as the curves in the spaghetti diagram (the
detailed shape of the spaghetti curves is unimportant), it immediately
suggests a successful model.  On the other hand, if the arrival-time
surface has unexpected minima or saddle points, and especially if the
unexpected features are far from identified lens images, that signals
an improbable model.  Fig.~\ref{fig:synth} shows what we call synthetic
images, meaning reconstructions of the extended lensed features by
fitting for a source.  These were generated by a new method, explained
in Appendix~\ref{subsec:sourcefit}, implemented during the offline
post-processing after the modelling process was complete.  The
synthetic images provided by SpaghettiLens during the collaborative
modelling and discussion were more crude; those are included in the
online supplement.

The arrival-time surface and synthetic image are summarised by one
diagnostic, the most important of all: are the lensed features
satisfactorily reproduced?  This diagnostic remains a judgment call by
modellers.  A useful quantitative criterion for whether the synthetic
image is consistent with the data would need to allow for PSF
dependence and unmodelled substructure -- otherwise all models would
be summarily rejected, something left for a future implementation
in SpaghettiLens.

\begin{figure*}
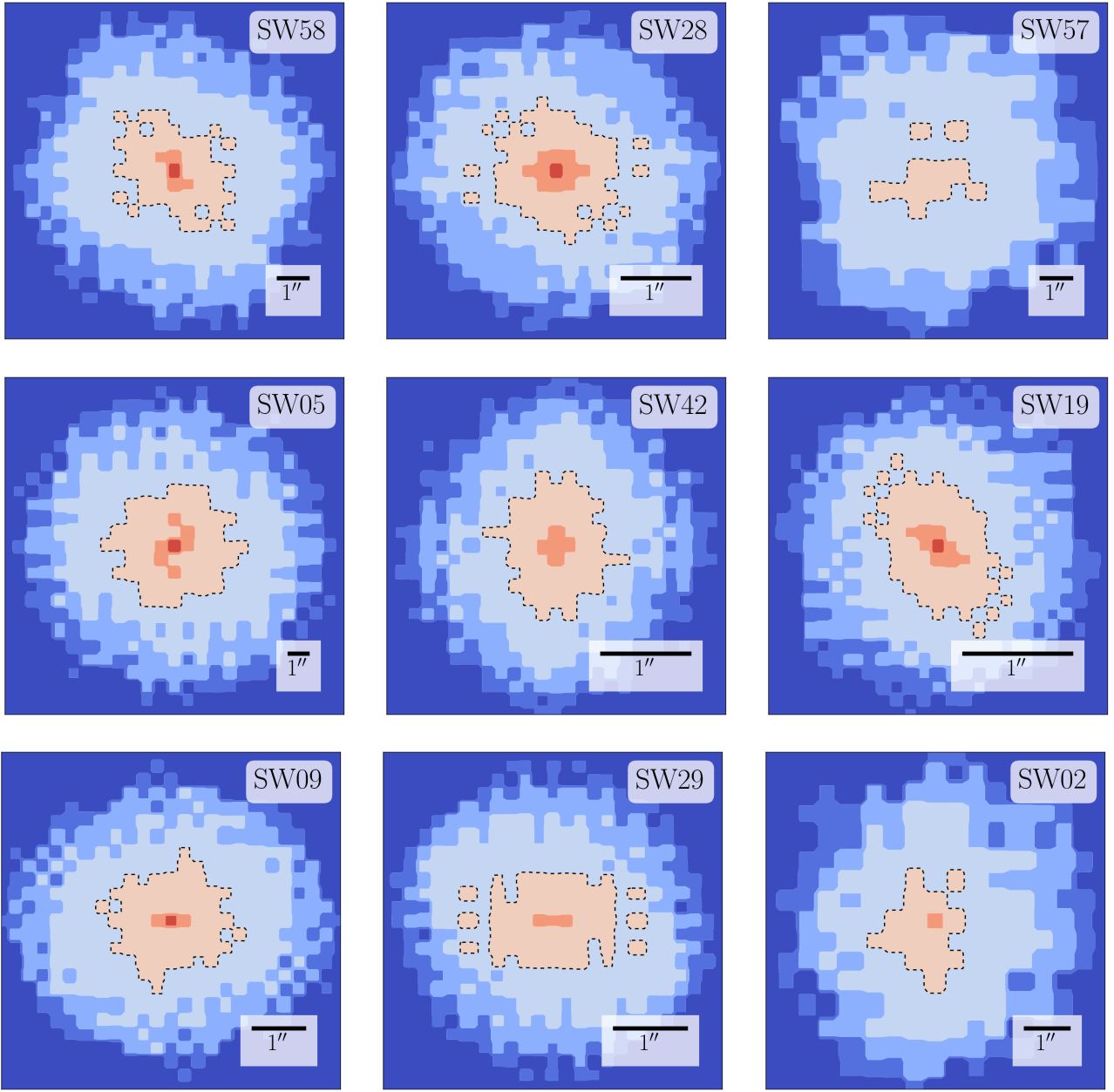

\inclGrid{kappa_map_interpol}
\caption{Ensemble-average mass distribution $\kappa$ for which the
  results Figures~\ref{fig:arriv}--\ref{fig:synth} were derived.  The
  dashed curves denote $\kappa=1$.  Most of the mass maps have a
  $180^\circ$-rotation symmetry, which is imposed by default.  For
  \sw{02} and \sw{57}, where the lensing mass is clearly asymmetric, the
  modeller chose to turn off the symmetry.
\label{fig:kappa}}
\end{figure*}

Fig.~\ref{fig:kappa} shows the projected mass maps of the sample.  In
fact, this figure only shows the ensemble-average mass maps, and not
the variation within the ensemble, from which uncertainties can be
inferred. The same applies to the arrival-time surfaces and synthetic
images in the previous two figures. The uncertainties will be shown in
a concise form in Fig.~\ref{fig:encl}.  Fig.~\ref{fig:kappa} makes
evident the tiled nature of the mass model.  The tiles can be smoothed
over by interpolation, and this was done in the mass maps during the
modelling process, available in the online version.  It is
interesting, however, to note the tiling artifacts, if only as a
reminder that the substructure in the mass distribution is very
uncertain, even if some integrated quantities are  well
constrained.  How the free-form mass maps relate to parameterised lens
models is discussed in Appendix~\ref{subsec:parameter}.
Note that although the mass distribution can have discontinuous jumps,
the lens equation and arrival-time surface are continuous.

Fig.~\ref{fig:encl} shows the enclosed-mass profiles, expressed as the
average convergence, $\kappa$, within circles of given projected
radius.  Uncertainties are included (see the figure caption for
details).
Appendix~\ref{subsec:hires} describes the improvements made
since our earlier work \citep{2015MNRAS.447.2170K}, to allow for
steeper profiles in the inner regions.  The enclosed mass is typically
best constrained at the notional Einstein radius, becoming more
uncertain at larger and smaller radii.

The mass maps and mass profiles are the basis of a further diagnostic:
are the mass distributions plausible? This is also a judgment call
made by modellers, but it showed to be a powerful diagnostic,
summarizing three aspects. The overall shape is forced to be
$180^\circ$-rotation symmetric, usually a plausible assumption, but
volunteers can deactivate this constraint. The profile slope turned
out to be a good indicator of plausible models, as can seen by
contrasting the model for \sw{57} with the rest of the sample:  The missing
core in Fig.~\ref{fig:kappa} and the flat profile in
Fig.~\ref{fig:encl} disqualify this model. The clumpiness of the
mass map is another useful indicator. Flat profile slopes can often be
identified directly in the mass map, where the mass tiles form a
checkerboard pattern.

More experienced volunteers applied these diagnostics already during the 
process of creating models as a criteria of a successful model, the evaluation 
presented in this work however was generated by the authors during post 
processing.

%%%%%%%%%%%%%%%%%%%%%%%%%%%%%%%%%%%%%%%%%%%%%%%%%%%%%%%%%%%%%%%%%%%%%%%
\section{Stellar and halo mass estimates}
\label{sec:stellar-mass}

The stellar masses of the lens galaxies are derived by comparison of
the photometric data with stellar $M/L$ estimates from population
synthesis models.  In principle, a detailed analysis of the spectral
energy distribution is needed to derive accurate stellar masses
\citep[e.g.][]{2009ApJS..185..253G,2011MNRAS.418.1587T}.  However,
at the redshifts probed, the photometric data mainly
constrain the information-rich 4000\thinspace\AA~break region, whose
strength depends sensitively on age and metallicity, thereby
providing a strong constraint on the stellar $M/L$.  Hence,
estimates to within $\sim$0.3\,dex in $\log(\Mstel/\Msun)$ can be derived
with a single colour, preferably tracing a rest-frame colour similar
to $U-V$ \citep[see Fig.~1 of][]{2008MNRAS.383..857F}, assuming
a universal initial mass function (IMF).  There is
evidence from detailed absorption line strength analysis that massive
galaxies can feature a non-standard IMF 
\citep[e.g.][]{2013MNRAS.429L..15F}. However, these variations --
towards a bottom-heavy distribution -- are typically found in the
cores of massive early-type galaxies \citep{2016MNRAS.457.1468L}. The
effect of these variations on the stellar $M/L$ of lensing systems is
still rather controversial
\citep{2015MNRAS.449.3441S,2016MNRAS.459.3677L}.

In this paper we further simplify the analysis by assuming a
relationship between the apparent total magnitude and stellar mass, at
the redshift of the lens.  For typical stellar-population parameters,
the variation of this relation is at most $\Delta\log$M$_s\sim$1\,dex.
A further potential systematic can arise from contamination of the
light of the lensing galaxy by the lensed background source.  Reducing
or eliminating the latter would require detailed fitting of light
distributions for each candidate \citep[see][]{2011ApJ...740...97L},
which we have not yet attempted.  Nonetheless, because the lensing
masses range over two orders of magnitude, it is still interesting to
compare them with rough estimates of stellar mass.

We make use of the \citet{2003MNRAS.344.1000B} models to derive two
functional forms of the stellar mass with respect to the $i^{\prime}$-band
magnitudes. The models have solar metallicity, with a Chabrier IMF,
and assume two different age trends: a ``young'' model, with a
constant 500\,Myr age at all redshifts, and an ``old'' model where the
age is the oldest one possible at each redshift, adopting a standard
$\Lambda$CDM model with $H_0=70$\,km\,s$^{-1}$\,Mpc$^{-1}$ and
$\Omega_m=0.3$.

Fig.~\ref{fig:stelmass} shows a comparison of stellar and lensing mass
in our sample.  The comparatively large span of the error bars in
stellar mass (horizontal axis) shows the range between the masses
derived using the two age trends respectively, and lies between 0.4 and
0.8\,dex.  It will be improved in future work by the use of available
optical and NIR magnitudes to derive more accurate constraints on the
stellar populations.  In addition, we also derive halo masses for the
lenses using an abundance-matching formula.  This technique matches
the distribution function of observed stellar mass in galaxies with
that of dark-halo masses from $N$-body simulations, to define  a simple
relation between stellar mass and halo mass.  We emphasize that a halo
mass from abundance matching should be considered an ``average''
estimate, and a significant scatter can be expected as galaxies with
the same stellar mass can be found in different environments. We refer
the reader to \cite{2012MNRAS.424..104L} for an assessment of the
effect of abundance matching on the derivation of dark matter halo
properties in lensing galaxies. We follow the prescription of
\citet{2010ApJ...710..903M}, namely:
\begin{equation}
    \frac{\Mstel}{\Mhalo} = \frac{2C_0}{(\Mhalo/M_1)^{-\beta} +
                                     (\Mhalo/M_1)^\gamma}
\end{equation}
\begin{align*}
    C_0 &= 0.02820, & M_1 &= 10^{11.884} M_\odot \\
    \beta &= 1.057, & \gamma &= 0.556.
\end{align*}
Fig.~\ref{fig:stelmass} may be compared with Fig.~4 in
\cite{2011ApJ...734...69M}.
The comparison of lensing and stellar mass produces the last 
of our model diagnostics, defined as a halo-matching index:
\begin{equation}
\haloindex \equiv \frac{\ln(M/\Mstel)}{\ln(\Mhalo/\Mstel)}
\end{equation}
that relates the observed lensing to stellar mass, with the
global ratio expected if the host halo corresponds to the
average value derived by abundance matching. Several cases
for $\haloindex$ can be considered:
\begin{itemize}
\item $\haloindex < 0$ is unphysical because $M<\Mstel$.
\item $\haloindex = 0$ means the stellar mass exactly accounts for the
  lensing mass (i.e. no dark matter affects the lensing model).
\item $0 < \haloindex < 1$ is the typical situation, where the lens
  includes stars and dark matter, but not the full halo.
\item $\haloindex = 1$ means that the lens consists of the entire halo.
\item $\haloindex > 1$ is in tension with abundance-matching, because the
  lensing mass exceeds the expected halo mass.
\end{itemize}
The halo-matching index expresses whether the lensing mass is
plausible given the flux received from the candidate lensing galaxy.

Fig.~\ref{fig:stelmass} and Table~\ref{tab:models} show that most of the
candidates have stellar
and lensing masses typical of massive ellipticals\footnote{The mass values
themselves are given in the online version of Table~\ref{tab:models}}.
\sw{05} is one of the most massive of all the candidates, corresponding
to a galaxy-group mass scale.  It is a particularly attractive system
for follow-up observations at higher resolution, as it is a large
system with clear multiple-image features. Modelling leaves little
doubt that it is a lens.  \sw{04} seems to be even more massive, but the
diagnostics leave some doubts about the validity of this model.  The
two lowest-mass systems, \sw{19} and \sw{42}, are important if they are
indeed lenses, as they would be low-mass lenses dominated by dark
matter.  All the modelled systems have reasonable stellar-mass
fractions, except for two cases where the stellar-mass fraction is too
low ($\haloindex > 1$): these are \sw{42} and \sw{57}.  In the case of \sw{57},
the model has poor diagnostics and should be discarded.  The model
for \sw{42}, on the other hand, is quite convincing -- except for the
high halo-matching index.  If \sw{42} turned out not to be a lens, that
would support the halo-matching index as an effective criterion to
discriminate models.

%%%%%%%%%%%%%%%%%%%%%%%%%%%%%%%%%%%%%%%%%%%%%%%%%%%%%%%%%%%%%%%%%%%%%%%
\section{Summary and conclusions}\label{sec:summary}

We report on a first set of mass distributions and follow-up
diagnostics for the {\SW} lens candidates created with a novel
approach that aims to be scalable by {\sl orders of magnitude} to
prepare for the many thousands of lenses the next generation of wide
field surveys will yield (e.g. Euclid, WFIRST).

Over the past few years, the way of discovering lenses has changed
with the introduction of machine learning and citizen science methods,
combined with the coverage of large areas of the sky by modern
surveys.  The way lensing mass models are constructed also needs to
change, in order to be prepared for the increasing influx of lens
candidates.  The work in this paper represents a hybrid approach
between the classical style -- where a small team of experts invest
many hours into the creation of a single model -- and a citizen
science project -- where a crowd of amateur volunteers make
independent contributions.  The authors of this paper are a
collaboration of professionals and experienced citizen-science
volunteers, aiming to create early-stage lens models as soon as a lens
candidate is found.

To assist volunteers in constraining lensing models, we introduce a
set of diagnostics that help asses the validity of the models. At a
later stage, we encourage modellers to apply those diagnostics as
feedback on the plausibility of their assumptions, and to suggest
additional diagnostics.

The diagnostics (i) -- (iv) (see Section.~\ref{sec:candidates_models})
turned out to be useful measures of the difficulty in modelling a
system, but they did not constitute necessary conditions for a
promising model.  They can help select systems to introduce novice
volunteers to the modelling process. In contrast, diagnostics (v) and
(vi) can be considered as necessary criteria for a good model.
Volunteers employed those ones to evaluate their models, and turned out
to be easy enough to grasp by new volunteers.  The halo-matching
index diagnostic (vii, $\haloindex$), is an interesting criterion that
might be useful for the modellers, but needs further investigation.

Table~\ref{tab:models} is a summary of our results.  It characterises
each modelled system with seven diagnostics, indicating (a)~the image
morphology and how clear or indistinct it is, (b)~whether the mass map
and synthetic lensed image appear to be plausible, and (c)~how the
model mass compares with the estimated stellar and full-halo masses.
Missing entries are due unavailable imaging data, whereas 
missing rows are due to models that were not created for this particular
candidate.

The trend in Fig.~\ref{fig:stelmass}, where higher-mass galaxies get
progressively more dark-matter dominated, is expected
\citep[see, e.g.][]{2005ApJ...623L...5F}, as is the span of about one
order of magnitude for the stellar mass and the two orders of
magnitude in lensing mass. With future data, it will be interesting to
compare the enclosed stellar and lensing mass as a function of radius,
going from the star-dominated inner regions to the outer dark
halos. \citet{2011ApJ...740...97L} illustrate this behaviour in their
Fig.~5, but the present sample could go an order of magnitude higher
in mass.

The quick creation of many models for the {\SW} candidates
successfully showed that a subset of citizen scientists are interested
in being involved in more challenging tasks that take some time to
learn. The next step involves recruiting more lensing enthusiasts, as
soon as the next round of {\SW} is started. In the meantime, the
improvements shown in the Appendix will be integrated in the standard
version of {\SpL}. Photometric fitting could also be
integrated into {\SpL}. This would allow experienced citizen
scientists to generate photometric redshifts and stellar masses, and
thus generate preliminary dark-matter maps as soon as a lens-candidate
is identified.

% \section{Acknowledgements}\label{sec:ack}
% We would like to thank the anonymous referee for comments
% that helped to improve this paper.

\bibliographystyle{template/mnras}
\bibliography{bib/bibli}

\clearpage

\begin{figure*}
\inclGrid{kappa_encl}
\caption{Cumulative circular-averages of the mass maps from
  Figure~\ref{fig:kappa}, with uncertainties.  More precisely, we show
  the enclosed mass within a given projected radius, expressed as the
  mean $\kappa$ with a given number of arcsec from the centre of the
  lensing galaxy.  The orange bands refer to the full ensemble of mass
  maps for the models, while the red curves show the ensemble
  averages.  The dashed vertical line indicates the notional Einstein
  radius, or where the mean enclosed $\kappa$ is unity.  The short
  vertical arrows marks the positions of the images (maxima, saddle
  points and minima).
  \label{fig:encl}}
\end{figure*}

\begin{figure*}
\includegraphics[width=\linewidth]{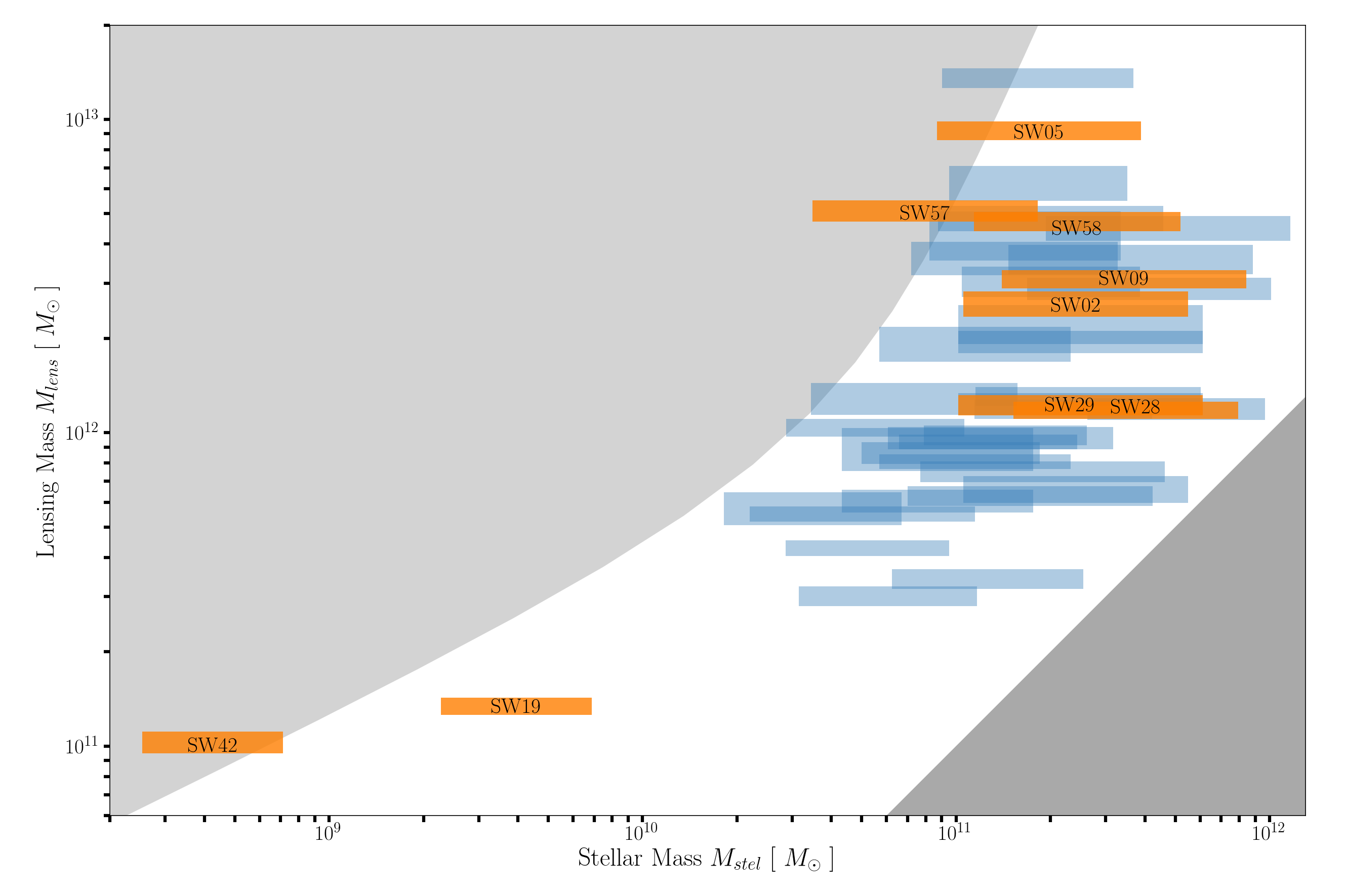}
\caption{Total mass in the model against the estimated stellar mass,
  alongside the values for the whole sample.  (The labelled orange
  bars are the systems shown in detail in
  Figures~\ref{fig:splinput}--\ref{fig:encl}.)  The horizontal extent
  of each bar indicates the extreme cases of a young (0.5~Gyr-old)
  stellar population and the oldest possible population at the given
  redshift.  The vertical extent indicates the spread of masses in the
  lens-model ensemble.  The lower-right shaded region is unphysical
  according to the stellar-population models, because it gives
  $M<\Mstel$. The upper-left shaded region is unphysical according to
  abundance matching (see Section~\ref{sec:stellar-mass}) because it
  gives $M>\Mhalo$.  That is to say, the unshaded region is
  $0<\haloindex<1$. \label{fig:stelmass}}
\end{figure*}

\begin{table*}
  \caption{Diagnostics of the selected models for each \SW
    candidate (see Section~\ref{sec:summary}).
    %A machine-readable version is available in the online supplement.
    }
  \label{tab:models}
  
\begin{tabular}{c c c | c | c c c | c | c c | c c c}
  \hline
  SWID & ZooID & CFHTLS Name
  
    & \rot{$z_\text{lens}$}

    & \multicolumn{1}{|l|}{\rot{\shortstack[l]{unblended\\images}}}
    & \rot{\shortstack[l]{all images\\discernible}}
    & \rot{\shortstack[l]{isolated\\lens}}

    & \rot{\shortstack[l]{image\\morpho-\\logy}}
    
    & \rot{\shortstack[l]{synthetic\\image\\reasonable}}
    & \rot{\shortstack[l]{mass map\\reasonable}}

    & \rot{\shortstack[l]{$\log_{10}\frac{\Mstel}{\Msun}$}\hskip-1.5pt}
    & \rot{\shortstack[l]{$\log_{10}\frac{M_\text{lens}}{\Msun}$}\hskip-2.2pt}
    & \rot{\shortstack[l]{halo-\\matching\\index $\haloindex$}}
  \\ \hline
 \sw{01} & \asw{4dv8} & J022409.5$-$105807 & \UK
    & \NO & \NO & \NO & LQ & \OK & \OK
    & \UK & \UK & \UK   \\
    
 \sw{02} & \asw{619d} & J140522.2$+$574333 & 0.7
    & \NO & \OK & \NO & SQ & \OK & \OK
    & 11.4 & 12.4 & 0.47   \\
    
 \sw{03} & \asw{6mea} & J142603.2$+$511421 & \UK
    & \OK & \NO & \NO & D & \OK & \OK
    & \UK & \UK & \UK   \\
    
 \sw{04} & \asw{9cjs} & J142934.2$+$562541 & 0.5
    & \OK & \NO & \NO & D & \NO & \OK
    & 11.3 & 13.1 & 0.93   \\
    
 \sw{05} & \asw{7k4r} & J143454.4$+$522850 & 0.58
    & \OK & \OK & \OK & IQ & \OK & \OK
    & 11.3 & 13.0 & 0.83   \\
    
 \sw{06} & \asw{8swn} & J143627.9$+$563832 & 0.5
    & \NO & \OK & \OK & SQ & \OK & \NO
    & 11.1 & 11.9 & 0.46   \\
    
 \sw{07} & \asw{7e08} & J220256.8$+$023432 & \UK
    & \OK & \OK & \NO & D & \OK & \OK
    & \UK & \UK & \UK   \\
    
 \sw{08} & \asw{99ed} & J020648.0$-$065639 & 0.8
    & \OK & \OK & \NO & D & \OK & \OK
    & 11.4 & 12.3 & 0.40   \\
    
 \sw{09} & \asw{2asp} & J020832.1$-$043315 & 1.0
    & \NO & \OK & \OK & SQ & \OK & \OK
    & 11.5 & 12.5 & 0.40   \\
    
 \sw{10} & \asw{2bmc} & J020848.2$-$042427 & 0.8
    & \OK & \NO & \OK & D & \NO & \NO
    & 11.3 & 11.9 & 0.29   \\
    
 \sw{11} & \asw{2qtn} & J020849.8$-$050429 & 0.8
    & \NO & \OK & \NO & LQ & \OK & \OK
    & 11.2 & 11.8 & 0.29   \\
    
 \sw{12} & \asw{3wsu} & J022406.1$-$062846 & 0.4
    & \OK & \OK & \NO & D & \OK & \OK
    & 10.8 & 11.5 & 0.44   \\
    
 \sw{13} & \asw{47ae} & J022805.6$-$051733 & 0.4
    & \NO & \NO & \NO & SQ & \NO & \NO
    & 11.1 & 11.9 & 0.46   \\
    
 \sw{14} & \asw{4xjk} & J023123.2$-$082535 & \UK
    & \NO & \NO & \NO & SQ & \NO & \OK
    & \UK & \UK & \UK   \\
    
 \sw{15} & \asw{4nan} & J084841.0$-$045237 & 0.3
    & \NO & \OK & \NO & CQ & \OK & \OK
    & 10.7 & 11.6 & 0.59   \\
    
 \sw{16} & \asw{9bp2} & J140030.2$+$574437 & 0.4
    & \NO & \NO & \OK & D & \NO & \OK
    & 11.3 & 12.1 & 0.34   \\
    
 \sw{17} & \asw{5rnb} & J140622.9$+$520942 & 0.7
    & \OK & \NO & \NO & D & \NO & \OK
    & 11.1 & 12.0 & 0.44   \\
    
 \sw{18} & \asw{7hu2} & J143658.1$+$533807 & 0.7
    & \OK & \NO & \OK & D & \NO & \NO
    & 11.4 & 12.1 & 0.31   \\
    
 \sw{19} & \asw{1ld7} & J020642.0$-$095157 & 0.2
    & \NO & \OK & \NO & IQ & \NO & \OK
    &  9.6 & 11.1 & 0.84   \\
    
 \sw{20} & \asw{2dx7} & J021221.1$-$105251 & 0.3
    & \OK & \OK & \OK & D & \NO & \OK
    & 11.2 & 12.0 & 0.44   \\
    
 \sw{21} & \asw{4m3x} & J022533.3$-$053204 & 0.5
    & \OK & \NO & \NO & D & \NO & \OK
    & 11.1 & 11.5 & 0.24   \\
    
 \sw{22} & \asw{9ab8} & J022716.4$-$105602 & 0.4
    & \NO & \NO & \NO & D & \NO & \OK
    & 11.7 & 12.1 & 0.15   \\
    
 \sw{23} & \asw{3r61} & J023008.6$-$054038 & 0.6
    & \NO & \OK & \NO & LQ & \NO & \OK
    & 11.2 & 12.6 & 0.71   \\
    
 \sw{24} & \asw{50sk} & J023315.2$-$042243 & 0.7
    & \NO & \OK & \NO & D & \OK & \OK
    & 11.4 & 11.8 & 0.19   \\
    
 \sw{25} & \asw{07mq} & J090308.2$-$043252 & \UK
    & \NO & \NO & \OK & D & \NO & \OK
    & \UK & \UK & \UK   \\
    
 \sw{26} & \asw{5ma2} & J135755.8$+$571722 & 0.8
    & \OK & \NO & \OK & D & \NO & \NO
    & 11.4 & 12.3 & 0.43   \\
    
 \sw{27} & \asw{6jh5} & J141432.9$+$534004 & 0.7
    & \NO & \NO & \NO & LQ & \NO & \OK
    & 10.7 & 11.7 & 0.67   \\
    
 \sw{28} & \asw{7xrs} & J143055.9$+$572431 & 0.7
    & \NO & \OK & \NO & LQ & \OK & \OK
    & 11.5 & 12.1 & 0.23   \\
    
 \sw{29} & \asw{8qsm} & J143838.1$+$572647 & 0.8
    & \NO & \OK & \OK & SQ & \OK & \OK
    & 11.4 & 12.1 & 0.31   \\
    
 \sw{30} & \asw{2p8y} & J021057.9$-$084450 & \UK
    & \OK & \NO & \NO & IQ & \NO & \NO
    & \UK & \UK & \UK   \\
    
 \sw{31} & \asw{21r0} & J021514.6$-$092440 & 0.7
    & \NO & \OK & \NO & LQ & \OK & \OK
    & 11.3 & 12.7 & 0.65   \\
    
 \sw{32} & \asw{4iye} & J022359.8$-$083651 & \UK
    & \NO & \OK & \NO & IQ & \OK & \OK
    & \UK & \UK & \UK   \\
    
 \sw{33} & \asw{3s0m} & J022745.2$-$062518 & 0.6
    & \OK & \OK & \NO & D & \NO & \OK
    & 10.9 & 12.1 & 0.77   \\
    
 \sw{34} & \asw{51ld} & J023453.5$-$093032 & 0.5
    & \NO & \NO & \OK & D & \NO & \OK
    & 10.9 & 11.9 & 0.59   \\
    
 \sw{35} & \asw{4wgd} & J084833.2$-$044051 & 0.8
    & \NO & \OK & \NO & LQ & \OK & \OK
    & 11.4 & 12.1 & 0.32   \\
    
 \sw{36} & \asw{096t} & J090248.4$-$010232 & 0.4
    & \OK & \OK & \NO & D & \NO & \OK
    & 11.0 & 12.0 & 0.56   \\
    
 \sw{37} & \asw{86xq} & J143100.2$+$564603 & \UK
    & \NO & \NO & \OK & SQ & \OK & \OK
    & \UK & \UK & \UK   \\
    
 \sw{38} & \asw{9cp0} & J143353.6$+$542310 & 0.8
    & \NO & \OK & \OK & LQ & \OK & \OK
    & 11.6 & 12.6 & 0.42   \\
    
 \sw{39} & \asw{5qiz} & J220215.2$+$012124 & \UK
    & \UK & \UK & \UK & \UK & \UK & \UK
    & \UK & \UK & \UK   \\
    
 \sw{40} & \asw{8wmr} & J221306.1$+$014708 & \UK
    & \NO & \OK & \OK & SQ & \OK & \OK
    & \UK & \UK & \UK   \\
    
 \sw{41} & \asw{8xbu} & J221519.7$+$005758 & 0.4
    & \OK & \NO & \OK & IQ & \OK & \OK
    & 10.5 & 11.8 & 0.80   \\
    
 \sw{42} & \asw{96rm} & J221716.5$+$015826 & 0.1
    & \OK & \OK & \NO & IQ & \OK & \OK
    &  8.6 & 11.0 & 1.04   \\
    
 \sw{43} & \asw{1c3j} & J020810.7$-$040220 & 1.0
    & \NO & \NO & \NO & SQ & \NO & \OK
    & 11.6 & 12.4 & 0.34   \\
    
 \sw{44} & \asw{2k40} & J021021.5$-$093415 & 0.4
    & \OK & \OK & \NO & LQ & \OK & \OK
    & 11.3 & 12.8 & 0.76   \\
    
 \sw{45} & \asw{24id} & J021225.2$-$085211 & 0.8
    & \NO & \OK & \OK & CQ & \NO & \OK
    & 11.7 & 12.6 & 0.37   \\
    
 \sw{46} & \asw{24q6} & J021317.6$-$084819 & 0.5
    & \OK & \OK & \NO & D & \OK & \OK
    & 10.9 & 11.8 & 0.49   \\
    
 \sw{47} & \asw{3r6c} & J022843.0$-$063316 & 0.5
    & \OK & \NO & \OK & D & \NO & \OK
    & 11.2 & 12.6 & 0.71   \\
    
 \sw{48} & \asw{0g95} & J090219.0$-$053923 & \UK
    & \OK & \NO & \OK & D & \OK & \OK
    & \UK & \UK & \UK   \\
    
 \sw{49} & \asw{07ls} & J090319.4$-$040146 & \UK
    & \NO & \OK & \OK & D & \OK & \OK
    & \UK & \UK & \UK   \\
    
 \sw{50} & \asw{08a0} & J090333.2$-$005829 & \UK
    & \OK & \NO & \OK & LQ & \OK & \OK
    & \UK & \UK & \UK   \\
    
 \sw{51} & \asw{6e0o} & J135724.8$+$561614 & \UK
    & \OK & \OK & \NO & D & \NO & \OK
    & \UK & \UK & \UK   \\
    
 \sw{52} & \asw{6a07} & J140027.9$+$541028 & \UK
    & \OK & \NO & \OK & SQ & \OK & \OK
    & \UK & \UK & \UK   \\
    
 \sw{53} & \asw{70vl} & J141518.9$+$513915 & 0.4
    & \OK & \NO & \OK & D & \NO & \OK
    & 11.3 & 12.5 & 0.56   \\
    
 \sw{54} & \asw{7sez} & J142620.8$+$561356 & 0.5
    & \NO & \OK & \NO & CQ & \OK & \OK
    & 11.1 & 12.3 & 0.68   \\
    
 \sw{55} & \asw{7t5y} & J142652.8$+$560001 & \UK
    & \NO & \OK & \OK & CQ & \OK & \NO
    & \UK & \UK & \UK   \\
    
 \sw{56} & \asw{7pga} & J142843.5$+$543713 & 0.4
    & \OK & \NO & \OK & D & \NO & \NO
    & 10.7 & 12.0 & 0.80   \\
    
 \sw{57} & \asw{8pag} & J143631.5$+$571131 & 0.7
    & \NO & \OK & \NO & SQ & \NO & \NO
    & 10.9 & 12.7 & 1.08   \\
    
 \sw{58} & \asw{7iwp} & J143651.6$+$530705 & 0.6
    & \NO & \NO & \OK & LQ & \OK & \OK
    & 11.4 & 12.6 & 0.58   \\
    
 \sw{59} & \asw{85cp} & J143950.6$+$544606 & \UK
    & \OK & \NO & \OK & D & \OK & \OK
    & \UK & \UK & \UK   \\

  \hline

\end{tabular}

\end{table*}

\clearpage

\appendix

%%%%%%%%%%%%%%%%%%%%%%%%%%%%%%%%%%%%%%%%%%%%%%%%%%%%%%%%%%%%%%%%%%%%%%%
\section{Developments in SpaghettiLens}

%%%%%%%%%%%%%%%%%%%%%%%%%%%%%%%%%%%%%%%%%%%%%%%%%%%%%%%%%%%%%%%%%%%%%%%
\subsection{Improved synthetic images}\label{subsec:sourcefit}

The mass maps produced by the current implementation of SpaghettiLens
are based on images of point-like features.  No information about
extended images is used, except in so far as they help the user
identify images of point-like features.  The synthetic images offered
to users are rudimentary, corresponding to conical light profiles
(i.e.  circular light profiles with brightness decreasing linearly
with radius).

\begin{figure}
  \includegraphics[width=\linewidth]{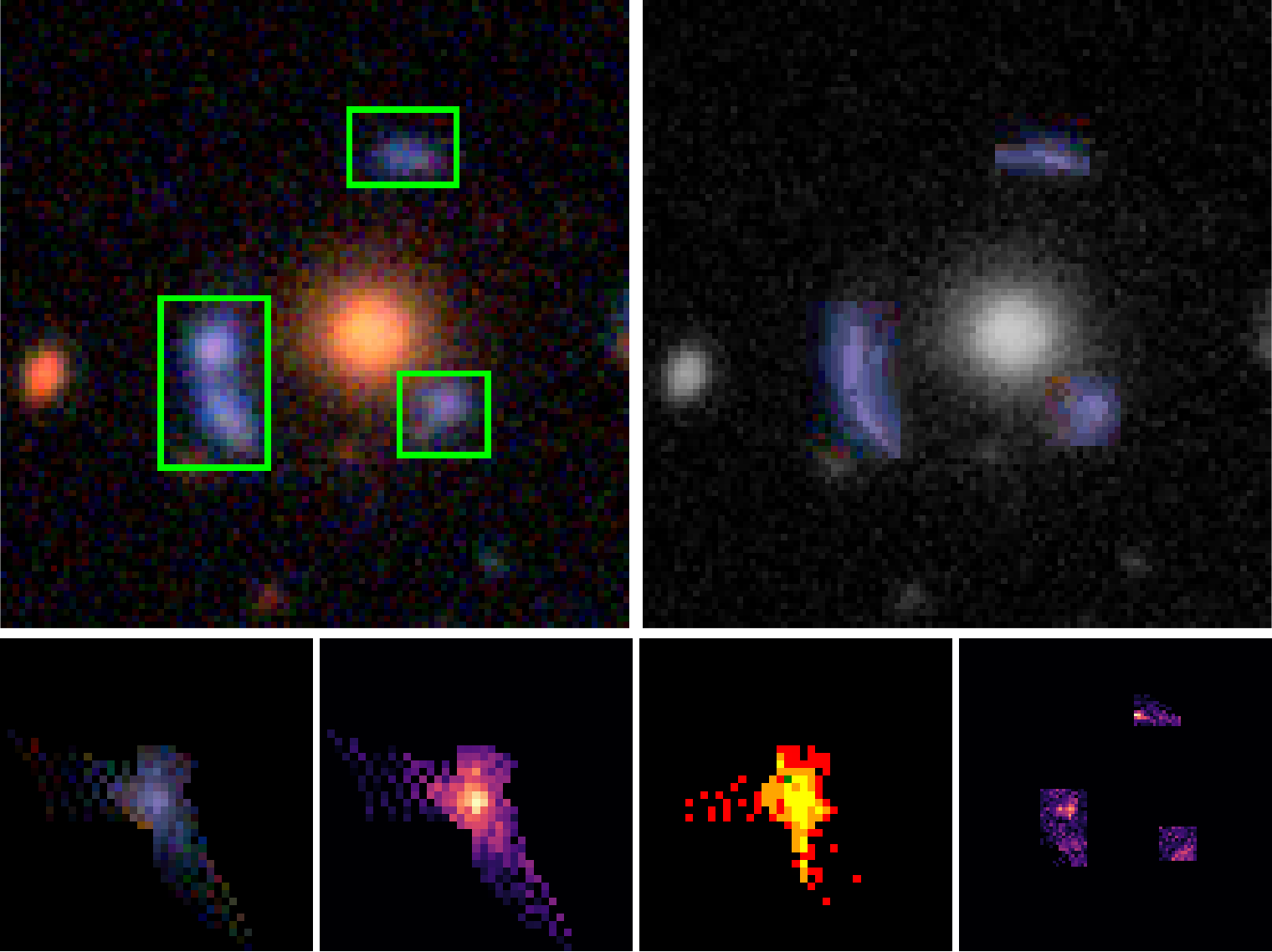}
  \caption{Synthetic lensed image with source-profile fitting in SW05
    (J143454.4+522850). Top-left: original image, with areas
    containing lensed images enclosed within green frames.  Top-right:
    synthetic image (coloured arcs) with lensing galaxy and unrelated
    objects in greyscale.  Bottom from left to right: reconstructed
    source in colour, intensity (greyscale), count of lens plane
    pixels per source plane pixel, residual of original image to
    synthetic image.}
  \label{fig:synthimg}
\end{figure}

We have now developed a prototype to improve the generation of
synthetic images, as illustrated in Fig.~\figref{synthimg}. Areas
containing lensed images are selected (green frames in the figure).
The selected areas should be as free as possible from other
sources, including the lensing galaxy. Pixels within the
selected areas are mapped to a grid on the source plane, using bending
angles given by the mass model.  The mapping from lens-plane pixels to
source-plane grid cells is many-to-one, because of image multiplicity
and magnification.  The brightness of each source-plane pixel is set
to the mean of all the lens-plane pixels mapped to it.  Finally, the
mapping is run back to the lens plane.  The result is a synthetic
image.  In effect, one is reconstructing a source-plane brightness map
by least-squares.

The procedure is not yet implemented in SpaghettiLens but can be
applied during post-processing.  The new synthetic images could be used to
improve the mass reconstruction, by weighing the ensemble of maps
according to how good the synthetic images are.

%%%%%%%%%%%%%%%%%%%%%%%%%%%%%%%%%%%%%%%%%%%%%%%%%%%%%%%%%%%%%%%%%%%%%%%
\subsection{Sub-sampling of the central region}\label{subsec:hires}

The models of simulated lenses in \cite{2015MNRAS.447.2170K} showed a
tendency to produce density profiles which were too shallow, resulting
typically in overestimates of the Einstein radius. Allowing some extra
mass tiles in central region, thus allowing the mass profile to rise
more steeply near the centre, was suggested as a possible cure.

\begin{figure}
\includegraphics[width=.9\linewidth]{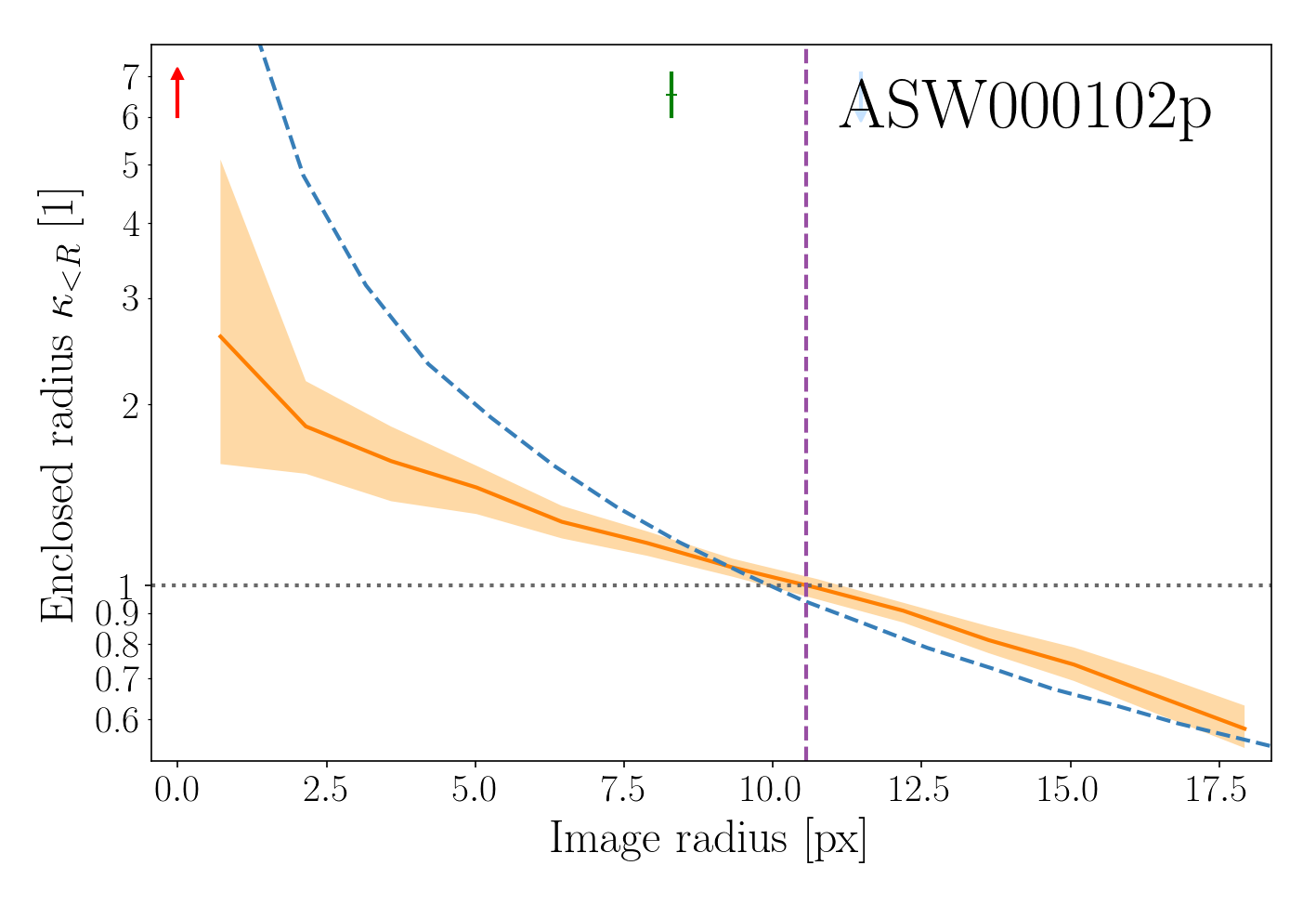}\\
\includegraphics[width=.9\linewidth]{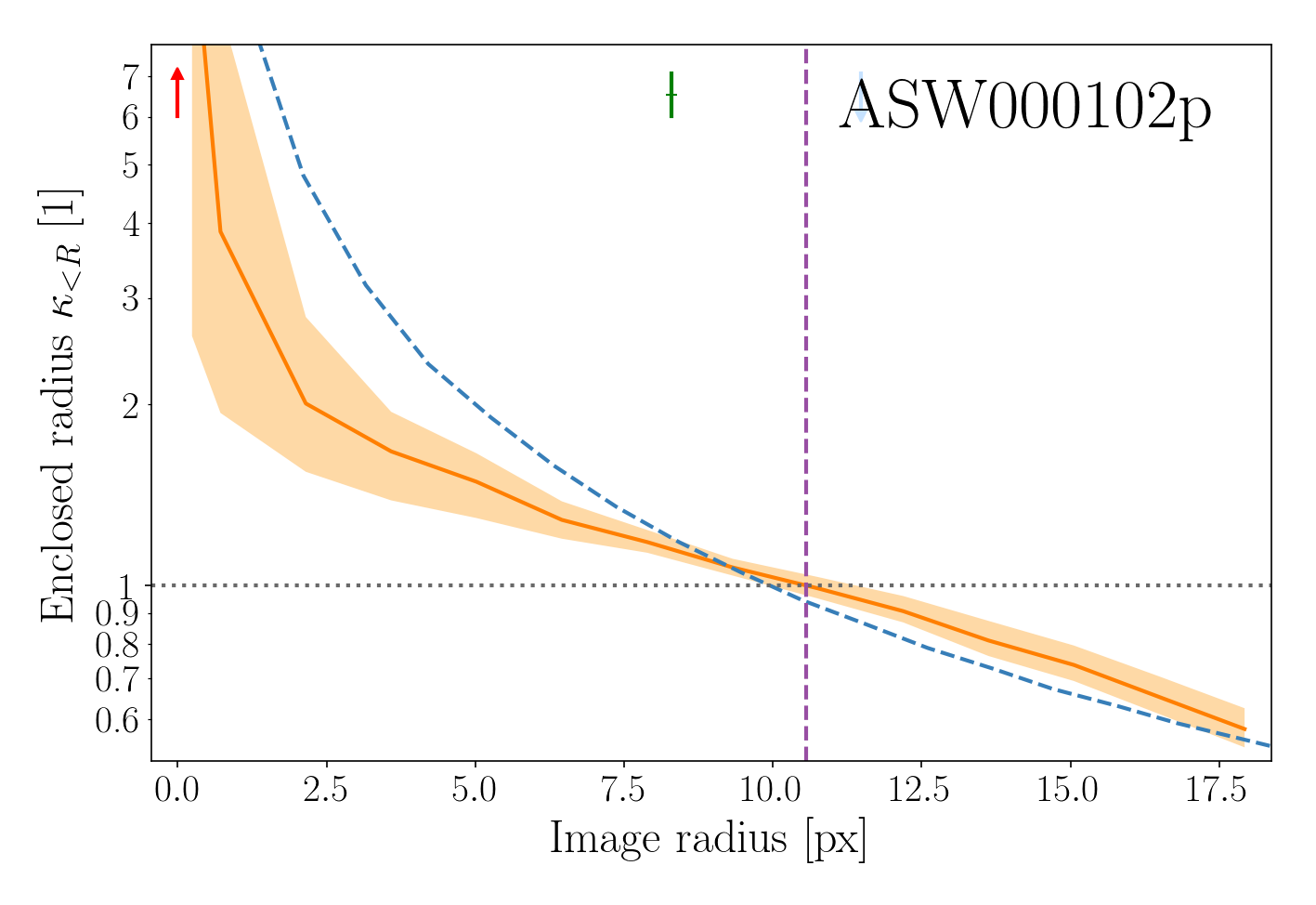}\\
\includegraphics[width=.9\linewidth]{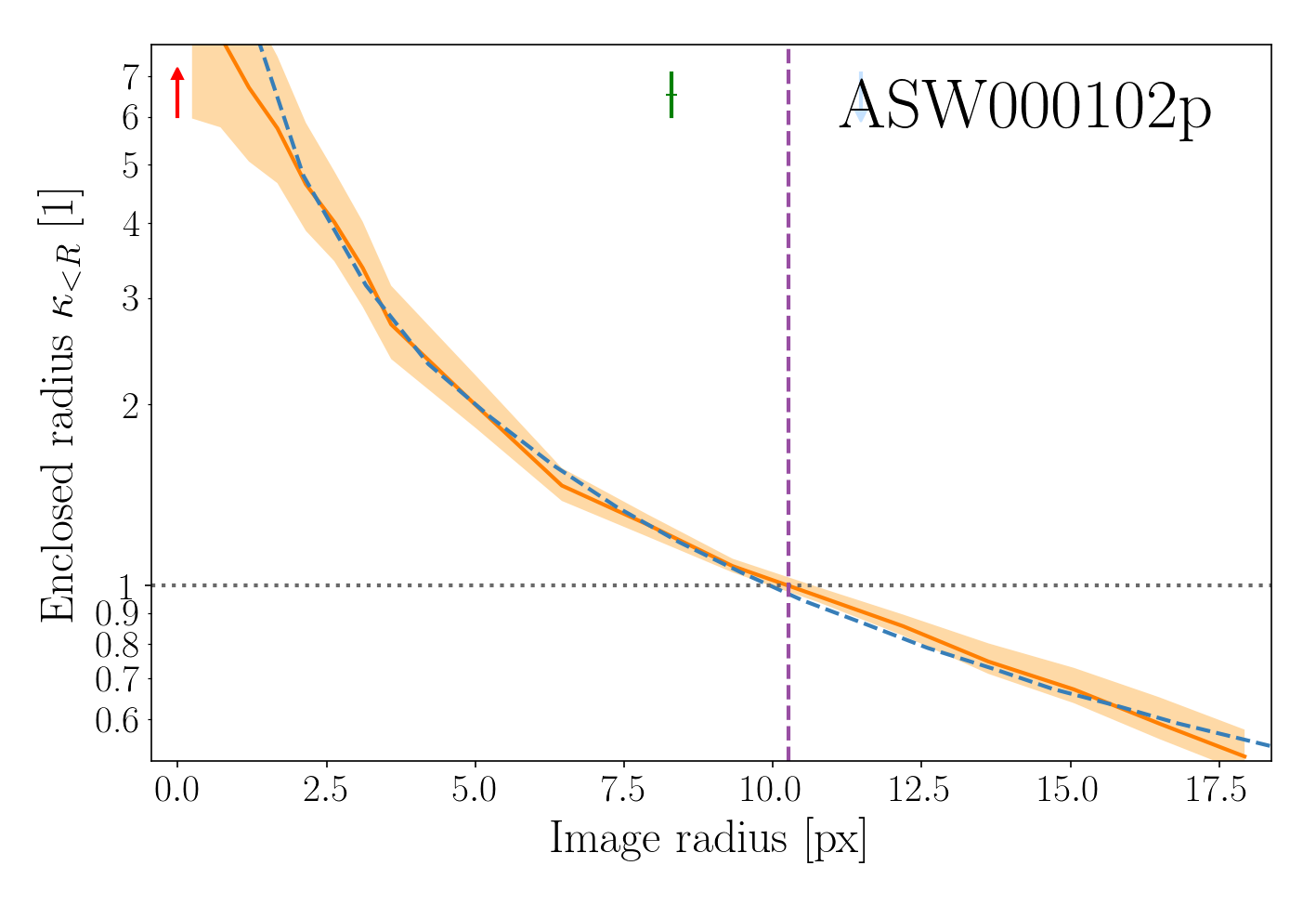}
  \caption{Model improvement resulting from the use of smaller mass tiles
    in the inner region of the mass model.  Shown here are the average
    enclosed $\kappa$ within a given projected radius, for three
    different reconstructions of a simulated lens (sim) from
    Space~Warps.  In each panel, the dashed blue curve is the correct
    answer.  The orange band represents the statistical ensemble from
    SpaghettiLens, the orange line shows the ensemble mean.  Locations of
    images (maximum, saddle point, minimum) are marked with vertical
    arrows.  The radial value at $\kappa=1$ is the effective
    Einstein radius, \ER. The upper panel is taken from
    \citet{2015MNRAS.447.2170K}, see Fig.~3 of that paper.  The middle
    panel is the result when the innermost mass tile is replaced by 9
    smaller tiles.  The lower panel results from replacing the
    innermost 5$\times$5 tiles with 9 smaller tiles each.}
  \label{fig:subsampling}
\end{figure}

Fig.~\figref{subsampling} shows an experiment with smaller mass
tiles in the inner region.  Replacing the very central mass tile with
9 smaller tiles allows for steeper central profiles.  Doing the same
for the 25 innermost mass tiles allows for even steeper central
profiles, eliminating the systematic shallow profiles.  However, 
it does not provide a completely satisfactory solution, because (a)~it increases
the number of mass tiles by 40\% and significantly increases the
computational time, and (b)~the square boundary between areas with
different tile sizes is rather undesirable.
The main modelling work in this paper was, however, done before the
experiments with smaller mass tiles was complete.  Some of the models
presented in this paper apply the intermediate option (corresponding
to the middle panel in Fig.~\figref{subsampling}) while others use
the old system.  We note this paper mainly concerns the enclosed mass
in the outer regions, so shallowness in the central region should not be
an issue.

%%%%%%%%%%%%%%%%%%%%%%%%%%%%%%%%%%%%%%%%%%%%%%%%%%%%%%%%%%%%%%%%%%%%%%%
\subsection{Parametrization of pixel models} \label{subsec:parameter}

\begin{figure}
  \includegraphics[width=\linewidth]{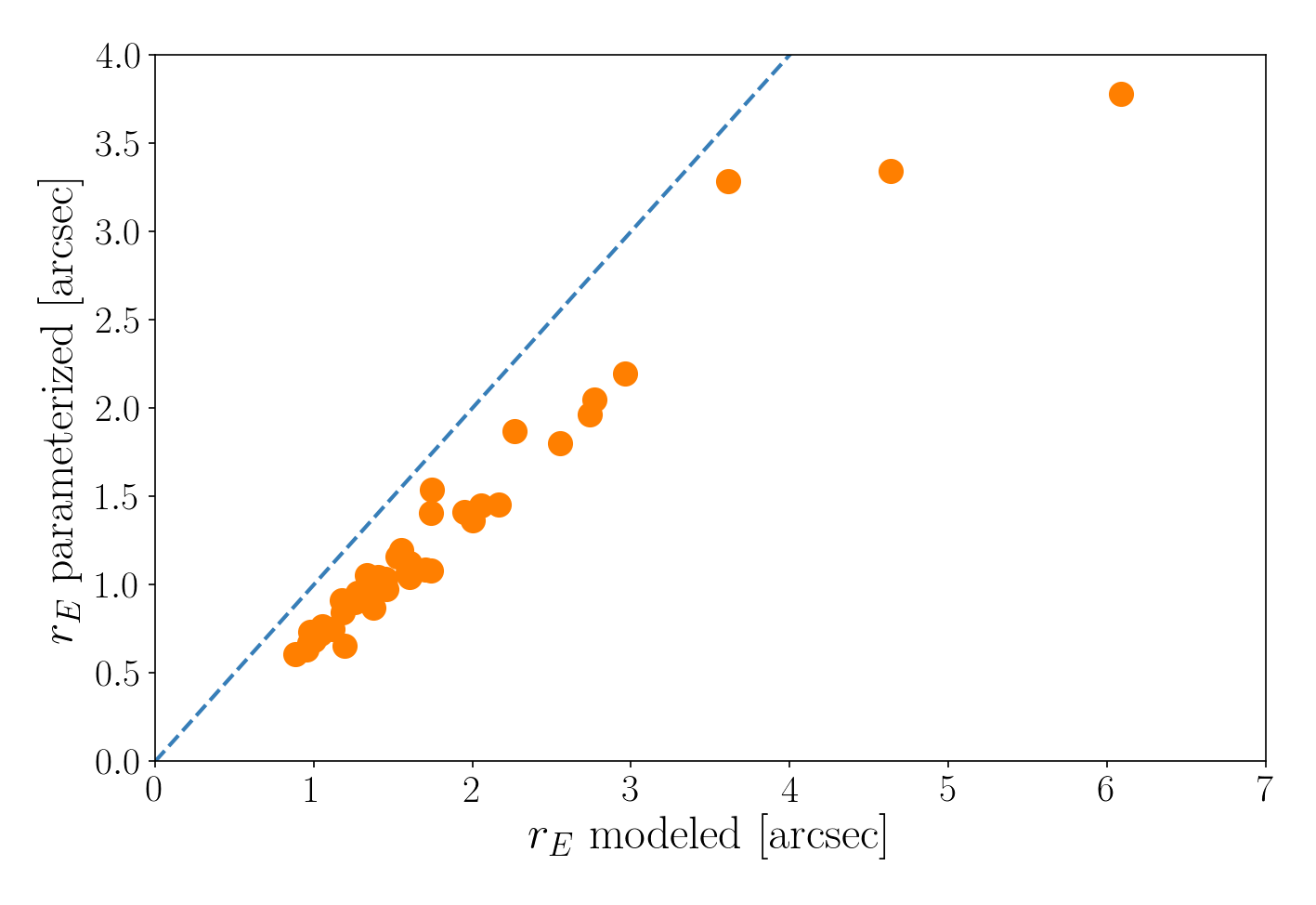}
  \caption{ Comparison of Einstein radii \ER\ obtained from mass tiles
    to the results from a parameterized model to test the performance
    of the algorithm.  The parameterized model was generated using
    principal component analysis on the ensemble of models.  The blue
    dashed line represents a perfect recovery of \ER.  }
  \label{fig:parameter}
\end{figure}

In order to fit the set of pixelated models to a single parametrized
model, a programme was written that took a parametrized function and
subtracted from it the mean and the principal components (PCs) of the
data. This created the residual function.  The number of principal
components in the analysis was varied, to test how this affected the
output. It was found that 5 PCs gave a reasonable approximation. A
masking function was added, selecting only the data points that fell
inside the image of the lens, and the PCs were clipped in order to
keep the values inside the region of the ensemble. Any value higher
than the clip was set to be the clip value. This was chosen to 2.5
because, assuming that the data follows a Gaussian error distribution,
almost all the values for the variance should lie between 2 and 3
standard deviations from the mean. Minimising the residuals function
produces the set of parameters that fit the parametrized function to
the original pixelated ensemble most closely.  A least squares fit was
used to perform this minimisation.  The parametrized model function
was obtained from the gravitational potential of an isothermal
ellipsoid mass distribution \citep{2001astro.ph..2341K}.  This model
is frequently used to describe gravitational lenses, giving good fits
to the observations.  The isothermal ellipsoid model outputs three
useful parameters: the radius of the Einstein ring, the ellipticity of
the model and the angle of the ellipticity from the vertical, giving
the orientation of the galaxy.  By applying this model to simulated
lenses for which the values of these parameters were already known, it
was possible to assess the accuracy of the methodology, before
applying the model to the candidate lensing galaxies.  Preliminary
results on the recovery of Einstein radii are shown in
Fig.~\ref{fig:parameter}.

% Don't change these lines
\bsp	% typesetting comment
\label{lastpage}
\end{document}